\documentstyle[epsf,12pt]{article}

\begin{document}

\title
{Universal invariant renormalization for supersymmetric
theories.}

\author{A.A.Slavnov\thanks{E-mail:$slavnov@mi.ras.ru$}\\
{\small{\em Steklov Mathematical Institute,
117966, Gubkina, 8, Moscow, Russia and}}\\
{\small{\em Moscow State University, physical faculty,
department of theoretical physics,}}\\
{\small{\em $117234$, Moscow, Russia}}
\\
\\
and K.V.Stepanyantz\thanks{E-mail:$stepan@theor.phys.msu.su$}\\
{\small{\em Moscow State University, physical faculty,
department of theoretical physics,}}\\
{\small{\em $117234$, Moscow, Russia}}}

\maketitle

\begin{abstract}
Manifestly invariant renormalization scheme for supersymmetric
gauge theories is proposed. This scheme is applied to supersymmetric
quantum electrodynamics.
\end{abstract}

\sloppy


\section{Introduction.}
\hspace{\parindent}

This paper addresses the problem of a manifestly invariant renormalization
procedure for supersymmetric (SUSY) gauge theories.

A peculiar feature of supersymmetric gauge theories is nonpolynomial
structure of the action. The only possibility to reduce the action to a
polynomial form is to use the Wess-Zumino gauge. However in this case the
manifest supersymmetry is lost and the powerful superdiagram technique
cannot be used.

In supersymmetric gauges an infinite number of primitively divergent
diagrams is present, which makes the theory formally nonrenormalizable.
Nevertheless it can be shown that supersymmetry and gauge invariance
reduce the number of independent counterterms, and the usual charge and
wave functions renormalizations are sufficient. The corresponding procedure
was firstly constructed for SUSY QED by one of the present authors
\cite{Sl1} and then applied to Abelian and non-Abelian models in papers
\cite{Sl2,FP,KSF}. However the proof given in these papers relies on the
assumption of a possibility to use an intermediate regularization preserving
the symmetry of the theory. Most popular gauge invariant regularizations
like dimensional regularization \cite{tHooftVeltman}, dimensional reduction
\cite{Siegel} or lattice regularizations break supersymmetry
\cite{JackJones}. Higher covariant derivative regularization
\cite{Slavnov,Bakeyev,West_Paper} in principle may be used, but the
calculations are quite involved (see however \cite{St1,St2}).

A possible alternative is presented by algebraic renormalization. In
this method one uses an arbitrary regularization or subtraction scheme,
breaking the symmetry of the theory (for example the momentum cut-off
regularization or the differential renormalization \cite{DiffR,Mas}).
The symmetry is restored at the second step by tuning finite counterterms
to provide relevant Generalized Ward Identities (GWI) for the renormalized
Green functions. This method was applied successfully to SUSY gauge
theories in the papers \cite{PS1,PS2,Mag}, where the invariant
renormalizability of $N=1$ and $N=2$ non-Abelian SUSY gauge models was
proven in the framework of algebraic renormalization. However a practical
implementation of the algebraic renormalization is rather cumbersome
as the procedure requires a tuning of a large number of (noninvariant)
counterterms.

Recently a new method of invariant renormalization was proposed
\cite{Sl3,Sl4}, which provides automatically the renormalized Green
functions possessing a relevant symmetry for arbitrary intermediate
regularization.  This method was formulated as a special subtraction
procedure which incorporates GWI. Solving explicitly the corresponding
GWI one reduces the problem of renormalization in an arbitrary
regularization scheme to explicitly gauge invariant procedure.
So instead of the two step algebraic renormalization we have the one
step procedure which guarantees the symmetry of the renormalized theory.

In the present paper we discuss how this method may be generalized to
supersymmetric gauge theories. Renormalization of SUSY QED is described
in details. A corresponding procedure for non-Abelian SUSY gauge models
is under consideration.

The paper is organized as follows:

In Section \ref{Section_SUSY_QED} we introduce the notations and
remind some information about supersymmetric quantum electrodynamics.
The universal invariant renomalization scheme for the model is constructed
in the next Section \ref{Section_Universal_Renormalization} and
illustrated by an example of the one-loop renormalization in Section
\ref{Section_One_Loop}. The results are discussed in the Conclusion.


\section{Supersymmetric quantum electrodynamics.}
\hspace{\parindent}
\label{Section_SUSY_QED}

In the superspace $N=1$ supersymmetric electrodynamics is described by
the following action:

\begin{eqnarray}\label{SQED_Action}
&& S_0 = \frac{1}{4 e^2} \mbox{Re}\int d^4x\,d^2\theta\,W_a C^{ab} W_b
+ \frac{1}{4}\int d^4x\, d^4\theta\,
\Big(\phi^+ e^{2V}\phi +\tilde\phi^+ e^{-2V}\tilde\phi\Big)
+\qquad\nonumber\\
&& + \frac{1}{2}\int d^4x\, d^2\theta\, m\, \tilde\phi\,\phi
+ \frac{1}{2}\int d^4x\, d^2\bar\theta\, m\, \tilde\phi^*\phi^*,
\end{eqnarray}

\noindent
where $\phi$ and $\tilde\phi$ are chiral superfields, $V$ is a real
superfield. The superfield $W_a$ in the abelian case is defined as

\begin{equation}
W_a = \frac{1}{16} \bar D (1-\gamma_5) D\Big[(1+\gamma_5)D_a V\Big],
\end{equation}

\noindent
where $D$ is the usual supersymmetric covariant derivative

\begin{equation}
D = \frac{\partial}{\partial\bar\theta} - i\gamma^\mu\theta\,\partial_\mu.
\end{equation}

\noindent
The integration over the superspace in the equation
(\ref{SQED_Action}) is defined as

\begin{eqnarray}
&& \int d^2 \theta = \frac{1}{4} \frac{\partial}{\partial \theta}
(1+\gamma_5) \frac{\partial}{\partial \bar \theta};\qquad
\int d^2 \bar \theta = \frac{1}{4} \frac{\partial}{\partial \theta}
(1-\gamma_5) \frac{\partial}{\partial \bar \theta};\nonumber\\
&& \int d^4 \theta = \int d^2\theta\,d^2\bar\theta
= \frac{1}{8} \Bigg(\frac{\partial}{\partial \theta}
\frac{\partial}{\partial \bar \theta} \Bigg)^2\vphantom{\int\limits^h}.
\end{eqnarray}

\noindent
Up to surface terms these expressions can be written in the explicitly
supersymmetric form:

\begin{eqnarray}
&& \int d^4x\, d^2 \theta
= - \frac{1}{4} \int d^4x\, \bar D (1+\gamma_5) D
= - \frac{1}{2} \int d^4x\, D^2;\nonumber\\
&& \int d^4x\, d^2 \bar \theta
= - \frac{1}{4} \int d^4x\, \bar D (1-\gamma_5) D
= - \frac{1}{2} \int d^4x\, \bar D^2;\nonumber\\
&& \int d^4x\, d^4 \theta = \frac{1}{4} \int d^4x\, \bar D^2 D^2
= \frac{1}{4} \int d^4x\, D^2 \bar D^2.\nonumber\\
\end{eqnarray}

\noindent
Here we use the following notations:

\begin{equation}
D^2 \equiv \frac{1}{2} \bar D (1+\gamma_5) D;\qquad
\bar D^2 \equiv \frac{1}{2} \bar D (1-\gamma_5) D.
\end{equation}

Action (\ref{SQED_Action}) is invariant under the gauge transformations

\begin{eqnarray}\label{Gauge_Transformation}
&& V \to V - \frac{1}{2} \Big(\Lambda + \Lambda^+\Big);\nonumber\\
&& \phi \to e^\Lambda \phi;\qquad \phi^* \to \phi^* e^{\Lambda^*};
\vphantom{\frac{1}{2}}\nonumber\\
&& \tilde\phi \to e^{-\Lambda} \tilde\phi;\qquad
\tilde \phi^* \to \tilde \phi^* e^{-\Lambda^*},\vphantom{\frac{1}{2}}
\end{eqnarray}

\noindent
where $\Lambda$ is an arbitrary chiral superfield. This invariance
allows to gauge away some components of $V(x,\theta)$, resulting in
the following equation:

\begin{eqnarray}\label{Wess_Zumino_Gauge}
V(x,\theta) = \frac{1}{2}\bar\theta \gamma^\mu \gamma_5\theta\, A_\mu(x)
+ i \sqrt{2} (\bar\theta\theta)(\bar\theta\gamma_5\chi(x))
+ \frac{1}{4} (\bar\theta\theta)^2 D(x),
\end{eqnarray}

\noindent
where $A_\mu$ is a gauge field, $\chi$ is a Maiorana spinor
and $D$ is a real scalar auxiliary field. In this gauge, which is
called Wess-Zumino gauge, residual gauge transformations depend only
on a single parameter, while the action is polynomial. However, this
gauge breaks explicitly supersymmetry of the model.

Quantization of model (\ref{SQED_Action}) is described in details
in book \cite{West} and is not considered here. We only note,
that the gauge fixing is made by adding the term

\begin{equation}\label{Gauge_Fixing}
S_{gf} = \frac{1}{32 e^2 \xi}\int d^4x\,d^4\theta\,
D^2 V\, \bar D^2 V,
\end{equation}

\noindent
$\xi$ being a constant.


\section{Universal invariant renormalization.}
\hspace{\parindent}
\label{Section_Universal_Renormalization}

In this section we consider a renormalization procedure, which may
be used with arbitrary not necessary gauge invariant regularization
providing renormalized correlators satisfying automatically SUSY GWI.
In the language of counterterms it means, that no noninvariant
counterterms are needed and renormalization freedom is restricted to
the choice of gauge invariant local terms, which are not fixed by GWI.
We assume, that a regularization used for calculations is manifestly
supersymmetric allowing to use supergraph technique. The construction
presented below may be used for arbitrary SUSY gauge models formulated
in terms of corresponding superfields. However in this paper we
concentrate on renormalization of SUSY QED and explicit equations will
refer to this model. We follow the approach, developed in \cite{Sl3,Sl4},
where the subtraction procedure incorporating automatically GWI was
proposed.

To avoid the appearance of spurious infrared divergences we shall
work in the "diagonal" gauge, corresponding to $\xi=-1$ in equation
(\ref{Gauge_Fixing}). Consideration of a general gauge requires additional
infrared regularization, which in the Abelian case may be achieved by
simply adding the mass term for gauge field and putting $m_A=0$ after
calculation of integrals (see \cite{Last_Reference}).

It is well known \cite{West}, that the degree of divergency of a
diagram with $E_\phi$ external lines of chiral and antichiral superfields
and $E_V$ external lines of the gauge superfield is equal to

\begin{equation}\label{Degree_Of_Divergence}
\omega = 2 - P - E_\phi,
\end{equation}

\noindent
where $P$ is a number of $\phi\phi$, $\tilde\phi\tilde\phi$,
$\phi^+\phi^+$ or $\tilde \phi^+ \tilde \phi^+$ propagators. Therefore,
divergencies are present in the following Green  functions:
\footnote{Note, that for an invariant regularization divergencies
in functions $\Pi$ will be absent for $n>2$. However, in the general
case it is impossible to ignore their existence.}

\begin{eqnarray}
&&
(2\pi)^4 \delta^4\Big(k_\mu + q_\mu + (p_1)_\mu + \ldots + (p_n)_\mu\Big)
\times\nonumber\\
&& \qquad\qquad\qquad \times
\Gamma\Big[(\theta_{x_1},p_1),\ldots (\theta_{x_n},p_n);(\theta_y,q),
(\theta_z,-q-p_1-\ldots-p_n)\Big]
\equiv \qquad\vphantom{\Bigg(}\nonumber\\
&& \equiv
\int d^4x_1\ldots d^4x_n\,d^4y\,d^4z\,
\frac{\delta^{n+2}\Gamma}{\delta V_{x_1}\ldots \delta V_{x_n}\,
\delta\phi_y\,\delta\phi^+_z}\Bigg|_{V,\phi,\tilde\phi=0}
\times\\
&& \qquad\qquad\qquad\qquad
\times \exp\Big(i (p_1)_\mu x_1^\mu + \ldots
+ i (p_n)_\mu x_n^\mu + i q_\mu y^\mu
+ i k_\mu z^\mu\Big).\nonumber\\
\nonumber\\
&&
(2\pi)^4 \delta^4\Big((p_1)_\mu + \ldots + (p_n)_\mu\Big)\,
\Pi \Big[(\theta_{x_1},p_1),\ldots
(\theta_{x_n},-p_1-\ldots-p_{n-1})\Big]
\equiv \qquad\vphantom{\Bigg(}\nonumber\\
&& \equiv
\int d^4x_1\ldots d^4x_n\,
\frac{\delta^{n}\Gamma}{\delta V_{x_1}\ldots \delta V_{x_n}}
\Bigg|_{\phi,\tilde\phi,V=0} \exp\Big(i (p_1)_\mu x_1^\mu + \ldots
+ i (p_n)_\mu x_n^\mu \Big)
\end{eqnarray}

\noindent
and in the functions $\tilde\Gamma$, which are constructed
similar to functions $\Gamma$, but differentiation is performed
over $\tilde\phi$-fields. Note, that functions $\Pi$ for odd $n$
are equal to 0, because contributions of diagrams with loops of
$\phi$ are cancelled with contributions of corresponding diagrams
with $\tilde\phi$-loops.

The functions $\Gamma$, $\tilde\Gamma$ and $\Pi$ satisfy Ward identities
\cite{KSF}, which in our notations can be written in the following form:

\begin{eqnarray}\label{Ward_Identity1}
&& (D^2_{x_1} + \bar D^2_{x_1})
\Gamma\Big[(\theta_{x_1},p_1),(\theta_{x_2},p_2),\ldots,(\theta_{x_n},p_n);
(\theta_y,q),(\theta_z,-q-p_1-\ldots-p_n)\Big]
=\vphantom{\frac{1}{2}}\nonumber\\
&&
= 2 \bar D^2_{x_1} \delta^4(\theta_y -\theta_{x_1})\times\nonumber\\
&& \qquad\qquad
\Gamma\Big[(\theta_{x_2},p_2),\ldots,(\theta_{x_n},p_n);(\theta_{x_1},q+p_1),
(\theta_z,-q-p_1-\ldots-p_n)\Big]
+\nonumber\\
&& + 2 D^2_{x_1} \delta^4(\theta_{x_1}-\theta_z)
\times\nonumber\\
&& \qquad\qquad
\times
\Gamma\Big[(\theta_{x_2},p_2),\ldots,(\theta_{x_n},p_n);(\theta_y,q),
(\theta_{x_1},-q-p_2-\ldots-p_n)\Big];
\\
\nonumber\\
\label{Ward_Identity3}
&& \Big(D^2_x+\bar D^2_x\Big)\Bigg(\Pi\Big[(\theta_x,p),(\theta_y,-p)\Big]
- \frac{1}{2 e^2\xi}\,p^2 \delta^4(\theta_x-\theta_y)\Bigg)=0;\\
\nonumber\\
&& \Big(D^2_{x_1} + \bar D^2_{x_1}\Big)\,\Pi \Big[(\theta_{x_1},p_1),
\ldots, (\theta_{x_{n-1}},p_{n-1}),
(\theta_{x_n},-p_1-\ldots-p_{n-1})\Big] = 0, \quad n>2.
\nonumber\\
\label{Ward_Identity4}
\end{eqnarray}

\noindent
The supersymmetric covariant derivative in momentum representation is
written as

\begin{equation}
D_x = \frac{\partial}{\partial\bar\theta} - \gamma^\mu p_\mu,
\end{equation}

\noindent
where $p$ is the momentum, corresponding to $x$-coordinate.
Ward identities for functions $\tilde\Gamma$ has the same structure.

As the chiral projector $E_c$ can be presented as

\begin{equation}
E_c \equiv \frac{1}{16\partial^2} \Big(\bar D^2 D^2 + D^2 \bar D^2\Big)
= \frac{1}{16\partial^2} \Big(\bar D^2 + D^2\Big)^2,
\end{equation}

\noindent
these identities express the Green functions with at
least one "chiral" gauge field component in terms of correlators with
less number of chiral gauge components.

Equations (\ref{Ward_Identity1}), (\ref{Ward_Identity3}) and
(\ref{Ward_Identity4}) are written for SUSY QED. However they have
essentially the same structure in non-abelian SUSY models, differing
by the RHS which includes in this case also correlators with Faddeev-Popov
ghost lines.

Our strategy may be formulated as follows: We firstly consider one-loop
diagrams and renormalize in arbitrary (infrared safe) way the diagrams
at the RHS of equations (\ref{Ward_Identity1}) -- (\ref{Ward_Identity4}).
Then having in mind that in an anomaly free theory GWI for correlators
calculated with arbitrary subtraction scheme may be violated only by
local terms, we define the renormalized Green functions at the LHS of
equations (\ref{Ward_Identity1}) -- (\ref{Ward_Identity4}) in such a way,
that these identities are satisfied automatically. This renormalization
still may be incomplete, as GWI obviously allow to add to the vertex
function at the LHS an arbitrary gauge invariant counterterm. These
counterterms as usual are free parameters which may be fixed by
normalization conditions.

In the case of SUSY QED this procedure looks as follows:

First of all it is necessary to renormalize one-loop two-point Green
function of matter superfields. The terms, proportional to $\phi^*\phi$
and not containing $V$ in the effective action, have the following
structure

\begin{equation}\label{Sigma_Definition}
4 \int d^4x\,d^4\theta\, \phi^*(x)
\Sigma\Big(\sqrt{-\partial^2}\Big) \phi(x).
\end{equation}

\noindent
Hence, the two-point function of the matter field can be written as

\begin{equation}\label{Two_Point_Function}
\Gamma\Big[(x,q),(y,-q)\Big]
= \bar D^2_x\, D^2_x \delta^4(\theta_x-\theta_y)\,\Sigma(q).
\end{equation}

\noindent
Renormalized two-point function is defined by subtraction

\begin{equation}\label{Sigma_Subtraction}
\Sigma^r(q) = \Sigma(q) - \Sigma(\mu_\sigma),
\end{equation}

\noindent
where $\mu_\sigma$ is a normalization point.

After renormalization of two-point function it is necessary to construct
renormalized vertex functions $\Gamma$ and $\tilde \Gamma$. Due to
supersymmetry the function $\Gamma$ can be presented in the form

\begin{equation}\label{GammaBF}
\Gamma[\theta,p] = \sum\limits_i B_i(\theta,p) F_i(p),
\end{equation}

\noindent
where $p$ and $\theta$ denote all set of arguments, $B_i(\theta,p)$
are polynomials in $p$ and $\theta$, which are some linear independent
combinations of covariant derivatives, acting on the product of
$\delta^4(\theta_k-\theta_l)$, and $F_i(p)$ are scalar functions of
external momenta. Renormalization is performed by subtracting
polynomials $P_i(p)$ from the functions $F_i(p)$. We choose these
polynomials in such a way, that the resulting function satisfies GWI
(\ref{Ward_Identity1}) -- (\ref{Ward_Identity4}) where the RHS includes
renormalized functions with less number of chiral external lines.

Let us define "partially renormalized" function

\begin{equation}\label{Partially_Renormalized_Function}
\gamma^r[\theta,p] = \sum\limits_i B_i(\theta,p) \Big(F_i(p)-P_i(p)\Big),
\end{equation}

\noindent
where $P_i(p)$ are some polynomials. Then the LHS of equation
(\ref{Ward_Identity1}) may be written in the form

\begin{equation}
(D^2_{x_1} + \bar D^2_{x_1}) \gamma^r[\theta,p]
= \sum\limits_i (D^2_{x_1} + \bar D^2_{x_1})
B_i(\theta,p) \Big(F_i(p)-P_i(p)\Big).
\end{equation}

\noindent
The combinations $\Big(D^2_{x_1}+\bar D^2_{x_1}\Big) B_i(\theta,p)$
in the general case are not independent. It is convenient to introduce
linear independent polynomials $Q_j$, proportional to the chiral parts of
$B_i$:

\begin{equation}
\Big(D^2_{x_1}+\bar D^2_{x_1}\Big) B_i(\theta,p)
= \sum_j c_{ij} Q_{j}(\theta,p).
\end{equation}

\noindent
Expanding the RHS of equation (\ref{Ward_Identity1}) over $Q_j$
it is possible to rewrite the Ward identity as

\begin{eqnarray}\label{Rewritten_GWI}
\sum\limits_{ij} c_{ij} Q_j(\theta,p) \Big(F_i(p)-P_i(p)\Big)
= \sum\limits_j Q_j(\theta,p) R_j(p).
\end{eqnarray}

\noindent
Taking into account, that $Q_j$ are linear independent, this equation
is equivalent to the following system of linear equations:

\begin{equation}\label{Eq}
\sum\limits_{i} c_{ij} \Big(F_i(p)-P_i(p)\Big) = R_j(p),
\end{equation}

\noindent
which represents Ward identity (\ref{Ward_Identity1})
expanded in terms of linear independent polynomials $Q_i(p,\theta)$.
If the polynomials $P_i(p)$ satisfy system (\ref{Eq}), the function
$\gamma^r$ defined by equation (\ref{Partially_Renormalized_Function})
satisfies SUSY GWI. Indeed,

\begin{equation}
\Big(D^2_{x_1}+\bar D^2_{x_1}\Big) \gamma^r[\theta,p]
= \sum\limits_i c_{ij} Q_j(\theta,p) \Big(F_i(p)-P_i(p)\Big)
= \sum\limits_j Q_j(\theta,p) R_j(p).
\end{equation}

\noindent
Note, that a choice of $P_i$ is not unique, because any solution of
Ward identity is defined up to a term $\Pi_{1/2} f$, which can not be
determined from the Ward identity. However, this freedom is irrelevant
for our procedure.

To eliminate the remaining ultraviolet divergences it is sufficient to
subtract from $\gamma^r$ gauge invariant local counterterms
$P_{gi}$

\begin{equation}
\Gamma^r[p,\theta] = \gamma^r[p,\theta]-P_{gi}.
\end{equation}

\noindent
These counterterms obviously can not be fixed by GWI. So we succeeded
to reduce the subtraction procedure in an arbitrary regularization
scheme to subtraction of gauge invariant counterterms.

Having obtained the function $\Gamma_3^r$, it is necessary to
substitute it into the RHS of renormalized Ward identity
(\ref{Ward_Identity1}) for the function $\Gamma_4^r$. Then the process
is repeated. So, we constructed a recurrent procedure, which defines
all functions $\Gamma_n^r$. This procedure is illustrated in the next
Section by an example of one-loop renormalization of the function
$\Gamma_3$ in the momentum cut-off regularization.

The functions $\tilde\Gamma^r$ are constructed in a similar manner.

At the next step it is necessary to renormalize Green functions $\Pi$,
corresponding to diagrams without external lines of matter superfields.
Due to supersymmetry they can be written in the form, similar to
equation (\ref{GammaBF}):

\begin{equation}\label{Pi_Structure}
\Pi[\theta,p] = \sum\limits_i B_i(\theta,p) F_i(p).
\end{equation}

\noindent
For example, the function $\Pi$ for $n=2$ can be presented as

\begin{eqnarray}\label{Original_Pi2}
&& \Pi\Big[(\theta_x,p),(\theta_y,-p)\Big]
= F_1(p)\,p^2\,\Pi_{1/2}\,\delta^4(\theta_x-\theta_y)
+ F_2(p)\,\delta^4(\theta_x-\theta_y),\qquad
\end{eqnarray}

\noindent
where

\begin{equation}
\Pi_{1/2} = - \frac{1}{16 \partial^2} D^a \bar D^2 C_{ab} D^b.
\end{equation}

\noindent
In this case

\begin{eqnarray}
B_1(\theta,p) = p^2 \Pi_{1/2}\,\delta^4(\theta_x-\theta_y);\qquad
B_2(\theta,p) = \delta^4(\theta_x-\theta_y)
\end{eqnarray}

As before, the renormalized Green functions are obtained by subtracting
from $F_i(p)$ some polynomials chosen to provide GWI for the function
$\Pi^r$

\begin{equation}\label{Renormalized_Pi}
\Pi^r[\theta,p] = \sum\limits_i B_i(\theta,p) \Big(F_i(p) - P_i(p)\Big),
\end{equation}

\noindent
For the two-point Green function substitution of
(\ref{Original_Pi2}) into equation (\ref{Ward_Identity3}) gives the
following equation:

\begin{equation}
\Big(F_2(p) - P_2(p) - \frac{1}{2 e^2\xi}\,p^2\Big)
\Big(D^2_x+\bar D^2_x\Big)\delta^4(\theta_x-\theta_y)=0.
\end{equation}

\noindent
Therefore,

\begin{equation}
P_2(p) = F_2(p) - \frac{1}{2 e^2\xi}\,p^2,
\end{equation}

\noindent
while the function $P_1$ can not be defined from GWI and corresponds to
a gauge invariant counterterm. It is convenient to choose

\begin{equation}
P_1(p) = F_1(\mu_\pi) - \frac{1}{2 e^2\xi}\,p^2,
\end{equation}

\noindent
where $\mu_\pi$ is a normalization point, which can be
different from $\mu_\sigma$. Then the renormalized two-point Green
function can be written as

\begin{eqnarray}\label{Pi2_Subtraction}
&& \Pi^r\Big[(\theta_x,p),(\theta_y,-p)\Big] =
\frac{1}{32 e^2 \xi}\Big(D^2 \bar D^2 + \bar D^2 D^2\Big)
\delta^4(\theta_x-\theta_y)
+\nonumber\\
&& \qquad\qquad\qquad\qquad\qquad\qquad\qquad
+ \Big(F_1(p) - F_1(\mu_\pi)\Big)\,p^2\,\Pi_{1/2}\,
\delta^4(\theta_x-\theta_y),\qquad
\end{eqnarray}

\noindent
This Green function satisfies the equation

\begin{equation}
\Big(D^2_x+\bar D^2_x\Big)\Bigg(\Pi^r\Big[(\theta_x,p),(\theta_y,-p)\Big]
- \frac{1}{2 e^2 \xi}\,p^2 \delta^4(\theta_x-\theta_y)\Bigg)=0,
\end{equation}

\noindent
which is a supersymmetric generalization of transversality condition
in the usual quantum electrodynamics.

Green functions, containing only external $V$-lines with $E_V > 2$,
can be renormalized similarly. It is sufficient to put in equations
(\ref{Rewritten_GWI}) and (\ref{Eq}) $R_j=0$. Then equation
(\ref{Partially_Renormalized_Function}) may be used to define the
renormalized function $\Pi^r$.

So, the one-loop renormalization procedure is finished. Due to the
locality of the subtraction terms in the limit, when regularization is
removed, the present scheme is equivalent to adding the counterterms

\begin{eqnarray}
&& \Delta S =
-\sum\limits_{n,i} \frac{1}{n!}
\int d^4x\,d^4\theta_1\,\ldots d^4\theta_n\,
B_i(\theta,\partial) P_i(\partial)\, V_1(\theta_1,x)\ldots V(\theta_n,x)
+\nonumber\\
&&
+ \frac{1}{4}\sum\limits_{n=0}^\infty (Z_{\Gamma_n}-1)\,
\frac{1}{n!} \int d^4x\,d^4\theta\,\phi^+ (2V)^{n} \phi
+\nonumber\\
&& \qquad\qquad\qquad\qquad\qquad\qquad
+ \frac{1}{4}\smash{\sum\limits_{n=0}^\infty} (\tilde Z_{\Gamma_n}-1)\,
\frac{1}{n!} \int d^4x\,d^4\theta\,\tilde \phi^+ (-2V)^{n} \tilde \phi.
\qquad
\end{eqnarray}

\noindent
It is important to note, that due to the presence of $\delta$-functions
in $B_i$ all these terms can be presented as integrals over a single
$\theta$. For the noninvariant regularization these counterterms certainly
can be not gauge invariant.

Having constructed one-loop counterterms, one can calculate
two-loop diagrams and perform similar renormalization at the two-loop
level. All combinatorics is given by the standard R-operation.
After the renormalization in the each loop renormalized Green functions
will automatically satisfy renormalized Ward identities. It means,
that the present scheme provides gauge invariant result for the effective
action even if a regularization is not gauge invariant.


\section{Application of universal invariant renormalization at
the one-loop level.}
\hspace{\parindent}
\label{Section_One_Loop}

To illustrate application of the scheme, described above, let us
consider one-loop renormalization of $N=1$ supersymmetric QED, regularized
by cutting-off loop momenta in the gauge with $\xi=-1$.

\begin{figure}[h]
\hspace*{4.3cm}
\epsfxsize7.0truecm\epsfbox{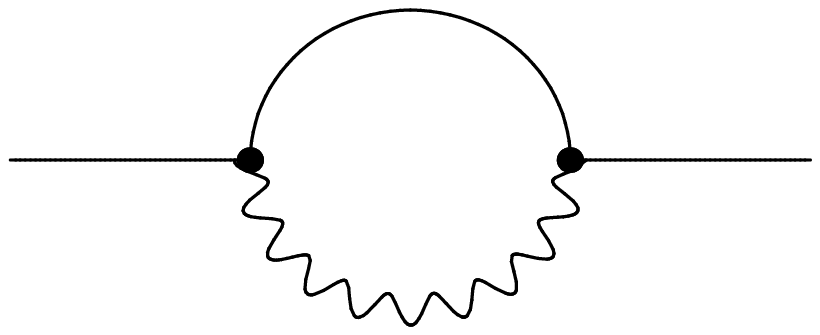}
\caption{Feinman diagram, giving nontrivial contribution
to the one-loop anomalous dimension.}
\label{Figure_Anomalous_Dimension_Diagram}
\end{figure}

The diagram, corresponding to the one-loop two-point Green
function of the matter field is presented at Figure
\ref{Figure_Anomalous_Dimension_Diagram}. After simple calculations,
using Feinman rules, described in book \cite{West}, in the Euclidean
space its contribution to the effective action can be written as

\begin{equation}
\quad \Delta\Gamma_\phi^{(1)} = -\int\frac{d^4q}{(2\pi)^4}\,
d^4\theta\,
e^2\,\Big(\phi^*(q) \phi(-q) + \tilde \phi^*(q) \phi(-q)\Big)
\int\limits^M \frac{d^4k}{(2\pi)^4}\,
\frac{1}{2 (k+q)^2 (k^2 + m^2)},\quad
\end{equation}

\noindent
where integration over loop momentum $k$ is defined as

\begin{equation}
\int\limits^M d^4k \equiv \int\limits_0^M dk\, k^3
\int\limits_0^\pi d\theta_1\,\sin^2\theta_1
\int\limits_0^\pi d\theta_2\,\sin\theta_2
\int\limits_0^{2\pi} d\theta_3
\end{equation}

\noindent
and $M$ is an ultraviolet cut-off. Therefore, according to equation
(\ref{Sigma_Definition}) the function $\Sigma(q)$ can be written as

\begin{eqnarray}
&& \Sigma(q) = -\int\limits^M \frac{d^4k}{(2\pi)^4}\,
\frac{e^2}{8 (k+q)^2 (k^2 + m^2)}
=\nonumber\\
&& \qquad\qquad\qquad\qquad
= -\frac{e^2}{128\pi^2} \Bigg(\ln \frac{M^2+m^2}{q^2+m^2}
+ 1 - \frac{m^2}{q^2} \ln\frac{q^2+m^2}{m^2} \Bigg).\qquad
\end{eqnarray}

Two-point function for the matter superfields is renormalized by
subtraction (\ref{Sigma_Subtraction}), which corresponds to

\begin{equation}
\Delta S = \frac{\alpha}{8\pi}
\Bigg(\ln \frac{M^2+m^2}{\mu_\sigma^2+m^2}
+ 1 - \frac{m^2}{\mu_\sigma^2} \ln\frac{\mu_\sigma^2+m^2}{m^2} \Bigg)
\int d^4x\, d^4\theta\,
\Big(\phi^* \phi + \tilde \phi^* \tilde\phi\Big).
\end{equation}

\begin{figure}[h]
\hspace*{1.5cm}
\epsfxsize12.0truecm\epsfbox{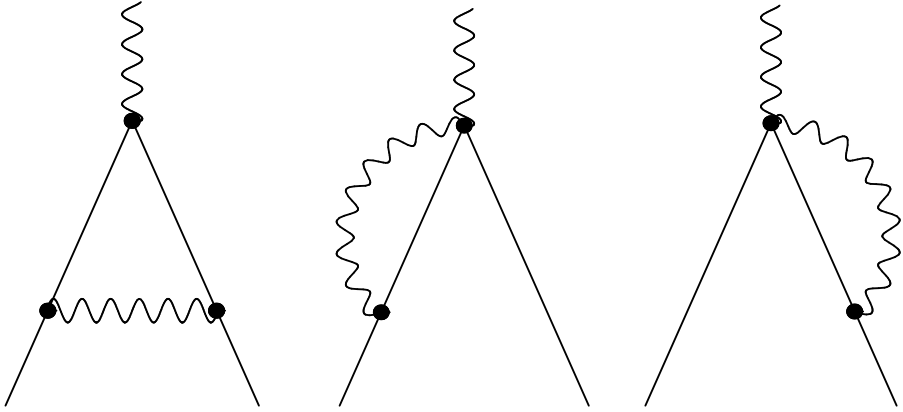}
\caption{Feinman diagrams, giving nontrivial contribution to the
one-loop three-point vertex function in the $N=1$ supersymmetric
electrodynamics.}
\label{Figure_Vertex_Diagrams}
\end{figure}

The next diagram to be considered is the one-loop three-point vertex
function, which is described by the diagrams, presented at Figure
\ref{Figure_Vertex_Diagrams}. Having calculated them in the chosen
regularization we obtained, that the corresponding three-point function
can be written as

\begin{eqnarray}\label{Gamma3}
&& \Gamma\Big[(\theta_x,p);(\theta_y,q),(\theta_z,-q-p)\Big]
=\nonumber\\
&&\qquad = \Bigg\{ -\int\limits^M \frac{d^4k}{(2\pi)^4}
\Bigg(\frac{e^2}{8 k^2 \Big((k+q)^2 + m^2\Big)}
+ \frac{e^2}{8 k^2 \Big((k+q+p)^2 + m^2\Big)}\Bigg)
-\nonumber\\
&&\qquad - \int\limits^M \frac{d^4k}{(2\pi)^4}
\frac{e^2 (k+q+p/2)_\mu}{4 k^2 \Big((k+q)^2 + m^2\Big)
\Big((k+q+p)^2 + m^2\Big)}\,\frac{1}{4}\bar D_x \gamma^\mu \gamma_5 D_x
+\nonumber\\
&&\qquad + \int\limits^M \frac{d^4k}{(2\pi)^4}
\frac{e^2}{8 k^2 \Big((k+q)^2 + m^2\Big)\Big((k+q+p)^2 + m^2\Big)}\,
p^2 (\Pi_{1/2})_x\Bigg\}
\times\nonumber\\
&& \qquad\qquad\qquad\qquad\qquad\qquad\qquad
\times
\Bigg(\bar D^2_x \delta^4(\theta_y-\theta_x)\,
D^2_x \delta^4(\theta_x-\theta_z)\Bigg).
\qquad\qquad
\vphantom{\Bigg(}
\end{eqnarray}

\noindent
The structure functions $B_i$ entering equation (\ref{GammaBF}) are

\begin{eqnarray}
&& B_1(\theta,p,q) = \bar D^2_x \delta^4(\theta_y-\theta_x)\,
D^2_x \delta^4(\theta_x-\theta_z);\vphantom{\frac{1}{2}}\nonumber\\
&& B_2(\theta,p,q) = \frac{1}{4}\bar D_x \gamma^\mu \gamma_5 D_x
\Big(\bar D^2_x \delta^4(\theta_y-\theta_x)\,
D^2_x \delta^4(\theta_x-\theta_z)\Big);\nonumber\\
&& B_3(\theta,p,q) = p^2 (\Pi_{1/2})_x
\Big(\bar D^2_x \delta^4(\theta_y-\theta_x)\,
D^2_x \delta^4(\theta_x-\theta_z)\Big).\vphantom{\frac{1}{2}}
\end{eqnarray}

\noindent
Renormalized vertex function is defined by subtracting the corresponding
polynomials:

\begin{eqnarray}\label{Gamma_R3}
&& \Gamma_3^r\Big[(\theta_x,p);(\theta_y,q),(\theta_z,-q-p)\Big]
=\nonumber\\
&& = \Bigg\{ -\int\limits^M \frac{d^4k}{(2\pi)^4}
\Bigg(\frac{e^2}{8 k^2 \Big((k+q)^2 + m^2\Big)}
+ \frac{e^2}{8 k^2 \Big((k+q+p)^2 + m^2\Big)}\Bigg) - P(p)
-\nonumber\\
&& -\int\limits^M \frac{d^4k}{(2\pi)^4}
\frac{e^2 (k+q+p/2)_\mu}{4 k^2 \Big((k+q)^2 + m^2\Big)
\Big((k+q+p)^2 + m^2\Big)}
\,\frac{1}{4}\bar D_x \gamma^\mu \gamma_5 D_x
+\nonumber\\
&& + \int\limits^M \frac{d^4k}{(2\pi)^4}
\frac{e^2}{8 k^2 \Big((k+q)^2 + m^2\Big)\Big((k+q+p)^2 + m^2\Big)}\,
p^2 (\Pi_{1/2})_x\Bigg\}
\times\nonumber\\
&& \qquad\qquad\qquad\qquad\qquad\qquad\qquad\qquad\qquad
\times
\Bigg(\bar D^2_x \delta^4(\theta_y-\theta_x)\,
D^2_x \delta^4(\theta_x-\theta_z)\Bigg).
\qquad
\vphantom{\Bigg(}
\end{eqnarray}

\noindent
Note, that we do not subtract anything from the last two terms, because
the corresponding integrals are convergent. Taking into account, that

\begin{eqnarray}
&& D^2\, \bar D \gamma_\mu \gamma_5 D = 4i D^2 \partial_\mu;
\nonumber\\
&& \bar D^2\, \bar D \gamma_\mu \gamma_5 D = - 4i \bar D^2 \partial_\mu;
\nonumber\\
&& \bar D^2 \Pi_{1/2} = D^2\Pi_{1/2} = 0,
\end{eqnarray}

\noindent
the left hand side of Ward identity can be written as

\begin{eqnarray}
&& \Big(D^2_x+\bar D^2_x\Big)\,
\Gamma_3^r\Big[(\theta_x,p);(\theta_y,q),(\theta_z,-q-p)\Big]
=\vphantom{\frac{1}{2}}\nonumber\\
&& = \bar D^2_{x} \delta^4(\theta_y -\theta_{x})
\bar D^2_{x} D^2_{x} \delta^4 (\theta_x-\theta_z)
\times\vphantom{\frac{1}{2}}\nonumber\\
&&\qquad\qquad\qquad
\times
\Bigg(-\int\limits^M \frac{d^4k}{(2\pi)^4}
\frac{e^2}{8 k^2 \Big((k+q+p)^2 + m^2\Big)}
- P(p)
\Bigg)+\qquad
\nonumber\\
&&
+ D^2_{x} \delta^4(\theta_{x}-\theta_z)
D^2_{x} \bar D^2_{x} \delta^4(\theta_x-\theta_y)
\times\vphantom{\frac{1}{2}}\nonumber\\
&&\qquad\qquad\qquad
\times
\Bigg(-\int\limits^M \frac{d^4k}{(2\pi)^4}
\frac{e^2}{8 k^2 \Big((k+q)^2 + m^2\Big)}
- P(p) \Bigg)
\end{eqnarray}

\noindent
while the right hand side of the Ward identity is written as

\begin{equation}
\quad 2 \bar D^2_{x} \delta^4(\theta_y -\theta_{x})
\bar D^2_{x} D^2_{x} \delta^4 (\theta_x-\theta_z)
\Sigma^r(q+p)
+ 2 D^2_{x} \delta^4(\theta_{x}-\theta_z)
D^2_{x} \bar D^2_{x} \delta^4(\theta_x-\theta_y) \Sigma^r(q).\quad
\end{equation}

\noindent
Thus, in this case

\begin{eqnarray}
&& Q_1(\theta,p,q) = \bar D^2_{x} \delta^4(\theta_y -\theta_{x})
\bar D^2_{x} D^2_{x} \delta^4 (\theta_x-\theta_z);\nonumber\\
&& Q_2(\theta,p,q) = D^2_{x} \delta^4(\theta_{x}-\theta_z)
D^2_{x} \bar D^2_{x} \delta^4(\theta_x-\theta_y).
\end{eqnarray}

\noindent
Taking into account, that

\begin{equation}
\Sigma^r(q) = -\int\limits^M \frac{d^4k}{(2\pi)^4}\,
\frac{e^2}{8 k^2 \Big((k+q)^2 + m^2\Big)} - \Sigma(\mu_\sigma)
\end{equation}

\noindent
we obtain, that

\begin{equation}
P(p) = \Sigma(\mu_\sigma).
\end{equation}

\noindent
Therefore, the renormalized three point function is equal to

\begin{eqnarray}\label{Gamma3_Renormalized}
&& \Gamma^r\Big[(\theta_x,p);(\theta_y,q),(\theta_z,-q-p)\Big]
= \Bigg\{ \Sigma^r(q) + \Sigma^r(q+p)
-\nonumber\\
&&\qquad - \int \frac{d^4k}{(2\pi)^4}
\frac{e^2 (k+q+p/2)_\mu}{4 k^2 \Big((k+q)^2 + m^2\Big)
\Big((k+q+p)^2 + m^2\Big)}\,\frac{1}{4}\bar D_x \gamma^\mu \gamma_5 D_x
+\nonumber\\
&&\qquad + \int \frac{d^4k}{(2\pi)^4}
\frac{e^2}{8 k^2 \Big((k+q)^2 + m^2\Big)\Big((k+q+p)^2 + m^2\Big)}\,
p^2 (\Pi_{1/2})_x\Bigg\}
\times\nonumber\\
&& \qquad\qquad\qquad\qquad\qquad\qquad\qquad\qquad
\times
\Bigg(\bar D^2_x \delta^4(\theta_y-\theta_x)\,
D^2_x \delta^4(\theta_x-\theta_z)\Bigg).
\qquad\qquad
\vphantom{\Bigg(}
\end{eqnarray}

It is evident, that our subtraction (together with the corresponding
subtraction for $\tilde\Gamma$) is equivalent to adding the following
counterterms:

\begin{equation}
\Delta S =  \frac{\alpha}{8\pi}
\Bigg(\ln \frac{M^2+m^2}{\mu_\sigma^2+m^2}
+ 1 - \frac{m^2}{\mu_\sigma^2} \ln\frac{\mu_\sigma^2+m^2}{m^2} \Bigg)
\int d^4x\, d^4\theta\,
\Big(\phi^+ 2V \phi - \tilde \phi^+ 2V \tilde\phi\Big),
\end{equation}

\noindent
where we substituted the explicit expression for $\Sigma(\mu_\sigma)$,
constructed above. So, the renormalization of the three-point function
is finished. At the next step it is necessary to consider four-point
function with two external matter superfields and continue the recurrent
process.

\begin{figure}[h]
\hspace*{1.3cm}
\epsfxsize12.5truecm\epsfbox{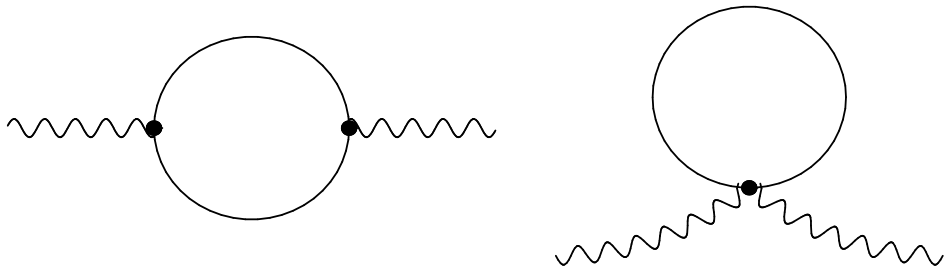}
\caption{Feinman diagrams, giving nontrivial contribution to the
one-loop $\beta$-function of $N=1$ supersymmetric electrodynamics.}
\label{Figure_Beta_Diagrams}
\end{figure}

The one-loop contribution to the two-point Green function of the gauge
field is defined by diagrams, presented at Figure
\ref{Figure_Beta_Diagrams} and is equal to

\begin{eqnarray}\label{One-loop_Beta_Contribution}
&& \int \frac{d^4p}{(2\pi)^4}\,
\Bigg\{\int
d^2\theta\, W_a(p) C_{ab} W_b(-p)
\int\limits^M \frac{d^4k}{(2\pi)^4}\,
\frac{1}{2 (k^2+m^2)\Big((k+p)^2 + m^2\Big)}
+\nonumber\\
&& + \int d^4\theta\, V(p)\,V(-p)
\int\limits^M\frac{d^4k}{(2\pi)^4}\,
\Bigg(\frac{1}{(k+p)^2+m^2} - \frac{1}{k^2+m^2}\Bigg)\Bigg\}.\qquad
\end{eqnarray}

\noindent
Note, that due to using the regularization, which breaks gauge
invariance in equation (\ref{One-loop_Beta_Contribution}) noninvariant
terms are present. Equation (\ref{One-loop_Beta_Contribution})
corresponds to

\begin{eqnarray}
&& F_1(p) = \int\limits^M \frac{d^4k}{(2\pi)^4}\,
\frac{1}{(k^2+m^2)\Big((k+p)^2 + m^2\Big)};\nonumber\\
&& F_2(p) = \int\limits^M \frac{d^4k}{(2\pi)^4}\,
\Bigg(\frac{2}{(k+p)^2+m^2} - \frac{2}{k^2+m^2}\Bigg)
= - \frac{p^2}{16\pi^2} + O(1/M^2).\qquad
\end{eqnarray}

\noindent
in equation (\ref{Original_Pi2}). (It is not difficult to
find explicit expression for $F_1(p)$, but it is rather
lengthy and we do not present it here.) After subtraction
(\ref{Pi2_Subtraction}) renormalized contribution of diagrams with two
external $V$-lines to the effective action can be presented as

\begin{eqnarray}\label{V2_Gamma_Contribution}
&& \frac{1}{2}\int \frac{d^4 p}{(2\pi)^4}\,d^2\theta\,
W_a(p) C_{ab} W_b(-p)\,\Big(F_1(p) - F_1(\mu_\pi)\Big),
\end{eqnarray}

\noindent
so all noninvariant terms disappear and the result becomes
finite. The $\beta$-function, corresponding to expression
(\ref{V2_Gamma_Contribution}), is equal to

\begin{equation}
\beta = \frac{\alpha^2}{\pi} + O(\alpha^3)
\end{equation}

\noindent
and agrees with calculations, made by dimensional reduction \cite{West}.

It is easy to verify, that the above subtraction is equivalent to
adding the following counterterms:

\begin{equation}
\Delta S = - \frac{1}{2} F_1(\mu_\pi)\,\mbox{Re}
\int d^4x\,d^2\theta\, W_a C_{ab} W_b
- \frac{1}{32\pi^2} \int d^4x\,d^4\theta\, V\,\partial^2 V,
\end{equation}

\noindent
which are not gauge invariant.


\section{Conclusion.}
\hspace{\parindent}

In this paper we presented a renormalization procedure for SUSY QED,
which guarantees gauge invariance of the renormalized theory for any
intermediate regularization. SUSY Ward identities are incorporated into
subtractions, which allows to avoid the appearance of noninvariant
counterterms.

The same procedure obviously may be applied to non-abelian supersymmetric
models provided they are formulated in terms of corresponding superfields.
Some technical problems related to more complicated structure of
non-abelian theories will be considered elsewhere.


\bigskip
\bigskip

\noindent
{\Large\bf Acknowledgments}

\bigskip

This work was partially supported by RBRF grant N 990100190 and by the
grant for support of leading scientific schools.


\pagebreak

\end{document}